\documentclass[lettersize,journal]{IEEEtran}
\usepackage{amsmath,amsfonts,mathabx}
\usepackage{algorithmic}
\usepackage{algorithm}
\usepackage{array}
\usepackage[caption=false,font=normalsize,labelfont=sf,textfont=sf]{subfig}
\usepackage{textcomp}
\usepackage{stfloats}
\usepackage{url}
\usepackage{verbatim}
\usepackage{graphicx}

\usepackage{multirow} 
\usepackage{nicefrac}       
\usepackage{siunitx} 
\usepackage{subfig}
\usepackage{svg} 
\usepackage{tabularx} 
\usepackage{booktabs, makecell} 
\usepackage{url}            
\usepackage{xcolor}
\usepackage[colorlinks=true, citecolor=black, linkcolor=black, urlcolor=black]{hyperref}
\usepackage{cleveref}
\crefname{figure}{Fig.}{Figs.} 
\usepackage[noadjust, nospace]{cite} 

\begin{document}

\title{Frequency noise characterization of diode lasers for vapor-cell clock applications}

\author{Gaspare Antona, Michele Gozzelino, Salvatore Micalizio, Claudio E. Calosso, \\Giovanni A. Costanzo, Filippo Levi
\thanks{G. Antona and G.A. Costanzo are with the Department of Electronics and Telecommunications, Politecnico di Torino, 10129 Torino, Italy (email: gaspare.antona@studenti.polito.it)}
\thanks{M. Gozzelino, S. Micalizio, C. E. Calosso,  and F. Levi are with the Quantum Metrology and Nanotechnologies Division, INRIM - Istituto Nazionale di Ricerca Metrologica, 10135 Torino, Italy (email: m.gozzelino@inrim.it)}}





\maketitle


\begin{abstract}
The knowledge of the frequency noise spectrum of a diode laser is of interest in several high-resolution experiments. Specifically, in laser-pumped vapor cell clocks, it is well established that the laser frequency noise plays a role in affecting clock performances. It is then relevant to characterize the frequency noise of a diode laser since such measurements are rarely found in the literature and hardly ever provided by vendors. In this paper, we describe a technique based on a frequency-to-voltage converter that transforms the laser frequency fluctuations into voltage fluctuations. In this way, it is possible to characterize the laser frequency noise power spectral density in a wide range of Fourier frequencies, as required in cell clock applications. 

\end{abstract}


\section{Introduction}

\IEEEPARstart{D}{iode} lasers are nowadays used in a large variety of scientific and technological applications. In atomic physics, in particular, they provide the necessary narrow-linewidth optical source to study the interaction of light and atoms \cite{wieman1991, camparo1985}. For this reason, diode lasers have increasingly become invaluable tools in several atomic physics sectors, including frequency standards \cite{matthey2011, ligeret2008, hermann2007, jiang2011}, laser cooling and trapping \cite{metcalf2007}, magnetometry \cite{budker2007}, and accurate gyroscopes \cite{gustavson1997, canuel2006}. 

In general, amplitude and spectral features of the laser light are of utmost importance for obtaining the best system performances. In fact, amplitude fluctuations, usually characterized by the relative intensity noise (RIN), represent a significant limitation in any optical measurement since they directly affect the signal-to-noise ratio. Analogously, laser phase/frequency fluctuations lead to a finite linewidth value, a parameter influencing the performance of high-resolution spectroscopy experiments \cite{maddaloni2013}.

Single-frequency laser diodes have recently become attractive also for generating low-phase-noise microwave radiation (optoelectronics oscillators) \cite{Maleki2017, Volyanskiy2010}. For this application, knowing the frequency-noise power spectral density (PSD) is necessary, since the frequency noise of the laser has a direct impact on the spectral purity of the generated microwave radiation.
\
%
\

\
In this paper, we will mainly address classes of lasers, such as distributed-Bragg-reflector (DBR) and distributed-feedback (DFB) lasers, 
used in vapor-cell frequency standards which represent a promising technology for near-future ground and space applications \cite{micalizio2021, godone2015, batori2021}.

Laser-pumped vapor-cell clocks have been realized following several operation schemes. More conventional devices work thanks to the double microwave-optical resonance approach where a microwave field excites the ground-state hyperfine transition in alkali vapors \cite{vanier2007}. A laser resonant with an optical transition is used to change the populations of the ground-state levels from their equilibrium value (optical pumping), increasing the number of interacting atoms. The clock reference transition is detected through the resulting change in absorption of the optical field as the frequency of the injected microwave is scanned across the ground-state transition.
In the coherent population trapping (CPT) approach \cite{vanier2005}, the clock resonance is excited by means of two phase-coherent laser fields separated in frequency by the ground-state hyperfine splitting. The absorption of the laser beams is used to monitor when the frequency difference between the lasers matches the clock frequency. More recently, double resonance and CPT approaches have been successfully operated also in pulsed regime \cite{godone2015}.

\begin{figure}[t]
\centering
\includegraphics[width=\columnwidth]{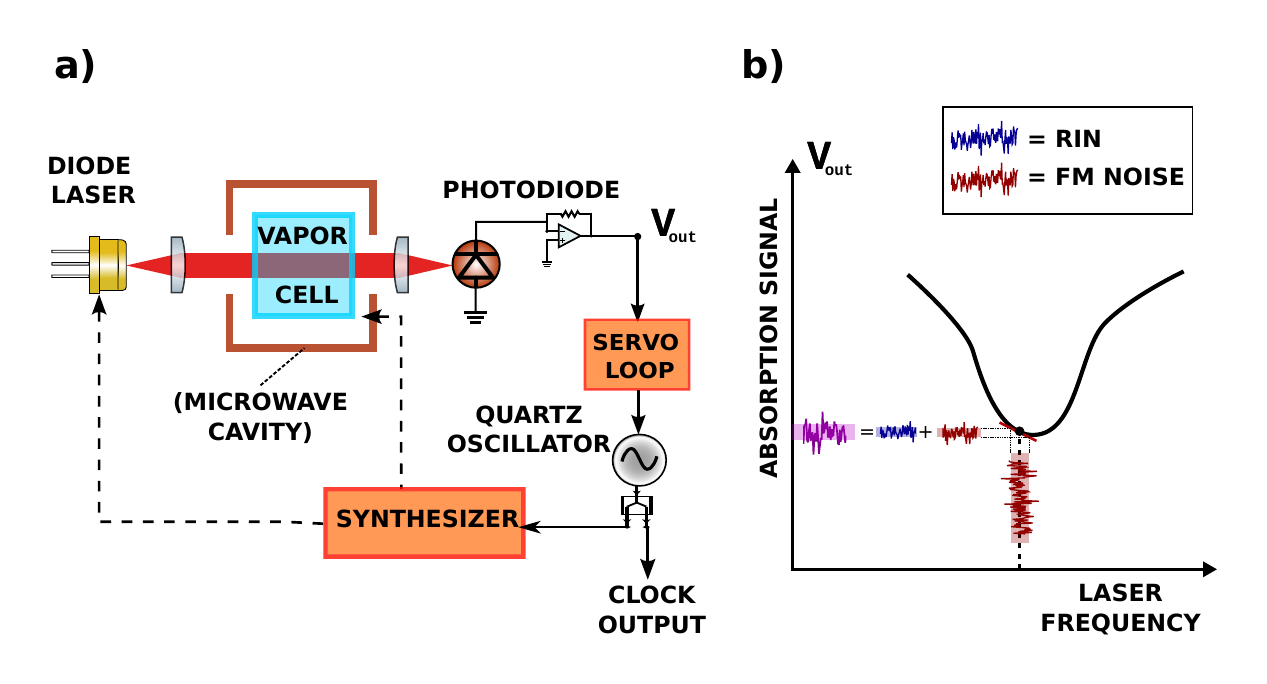}
\caption{Basic configuration of a typical cell clock. In the double resonance scheme, a diode laser is used to pump and probe the ground-state atomic population, whereas the clock transition is probed with a microwave field sustained by a microwave cavity. In the case of the CPT method, the microwave generated with the synthesizer is either used directly to modulate the laser frequency (creating two sidebands) or to phase-lock two lasers with a frequency offset equal to the synthesizer frequency. In either case, the clock signal is the result of an absorption measurement, with the laser propagating through the resonant vapor. b) Simplified sketch of the laser noise sources: amplitude (RIN) and frequency (FM). Both noise types affect the photodiode signal and enter the servo loop, limiting the clock stability.}
\label{fig:1}
\end{figure}

Regardless for the specific technique adopted to implement the clock, we can say that in all laser-pumped vapor cell frequency standards the clock resonance is detected by observing the light transmitted through the cell on a photo-diode, as shown in \cref{fig:1}a. In other words, the optical signal carries the information about a clock transition occurring in the microwave domain. As shown in \cref{fig:1}b, it is evident that laser instabilities can significantly affect the atomic signal of vapor-cell devices, as they employ an absorption measurement scheme. In particular, laser intensity noise is transmitted through the resonance cell and is detected by the photodetector, affecting the clock signal that is used to close the loop on the local oscillator (usually a quartz oscillator).On the other hand, due to optical pumping, laser frequency fluctuations are converted into amplitude fluctuations at the output of the cell, again affecting the clock signal \cite{Camparo1998,camparo1999,Kitching2001}. As a result, laser frequency noise is one of the critical aspects to consider when designing a cell clock \cite{almat2018_dr,yun2017,hafiz2017}.  


Very often, the information about the laser frequency fluctuations is simply expressed in terms of the laser line shape and its associated linewidth. However, for vapor cell applications, this information is not sufficient, and the complete knowledge of the frequency-noise PSD is needed. Indeed, the clock servo loop requires modulation of the interrogating signal, introducing a characteristic modulation frequency (for continuously operated clocks \cite{yun2017, bandi2014}) or a cycle frequency (for clocks working in pulsed operation \cite{micalizio2012, kang2015, hafiz2017}). Due to aliasing, noise at multiples of this modulation/cycle frequency will directly impact the clock stability. For clocks employing centimeter-scale cells, the modulation frequency is typically in a range of Fourier frequencies from \SI{100}{\hertz} to \SI{1}{\kilo\hertz}, depending on the specific implementation.
The knowledge of the laser frequency-noise PSD in this particular range of Fourier frequencies is therefore particularly relevant for choosing the correct laser source and for determining a complete clock stability budget.

Various methods exist for measuring the frequency noise of a laser. One of these is the homodyne detection or delay-line method: the laser light is split into two arms and one is delayed with respect to the other before recombining both beams. The resulting response is measured with a fast Fourier transform (FFT) spectrum analyzer \cite{llopis2011}. 

A variation of this technique is the self-heterodyne approach: one portion of the laser beam is sent through a long optical fiber which provides some time delay. The other portion is sent through an acoustic-optic modulator. The two beams are then superimposed on a beam-splitter and the frequency-shifted beatnote is detected by a photodiode. This technique rejects near-DC noise and a standard RF spectrum analyzer can be used to measure the frequency noise spectrum \cite{ludvigsen1994}. When the time delay introduced by the delay line in the interferometer is larger than the coherence time, the mixed signals will be uncorrelated and the measured beatnote signal is not sensitive to the mean phase difference of the fields. However, for correlated fields, the observed homodyne spectrum becomes critically dependent on the optical phases \cite{horak2006}. It is a convenient method when only one laser source is available.

\
Another method exploits an atomic resonance as a frequency discriminator by tuning the laser frequency on the side of an optical transition \cite{didomenico2010,tombez2012}. This technique is convenient as it does not requires a reference laser, but it applies only to laser sources resonant with specific atomic transitions. Moreover, the slope of the frequency discriminator needs to be determined before each measurement, as the laser detuning from resonance or the atomic lineshape may vary with the experimental conditions. In this scheme, laser amplitude fluctuations add linearly to the measured signal, possibly limiting the measurement accuracy, especially at high Fourier frequencies \cite{Bartalini2011}.

Finally, we mention the measurement scheme exploited in \cite{Volyanskiy2010}. There, a differential phase shift keying (DPSK) modulator is used as frequency-to-amplitude discriminator. As in the case of the atomic-resonance method, this scheme is not immune to amplitude fluctuations either coming from the laser source or from the detection noise.

We propose an alternative and robust method to measure the laser frequency spectrum. The method is based on the generation of an optical beatnote, translating the laser noise into the RF domain, and on the analysis of the beatnote frequency noise. The noise analysis is based on frequency-to-voltage (f/V) conversion \cite{cohen1973,yan2018,antona2022}. This approach is motivated by the fact that the phase noise of semiconductor laser diodes is typically too high to be characterized by high-performance commercial phase meters, usually designed to measure low-noise RF oscillators. This technique overcomes the limitation of the phase meter dynamics, providing the needed sensitivity for measuring the frequency noise of a wide class of diode lasers.

The paper is organized as follows: in \Cref{sec:convertitore} we present a detailed description of the measurement scheme and the characterization of the f/V converter in terms of sensitivity and noise. In \Cref{sec:comparison}, we compare the proposed method to the already mentioned measurement schemes, in terms of applicability and sensitivity.
In \Cref{sec:validation}, to validate the method, we apply it to the characterization of a distributed feedback laser, in terms of its frequency-noise PSD. 
Conclusions are reported in \Cref{sec:conclusioni}.
         
\section{Frequency-noise measurement scheme}
\label{sec:convertitore}
The proposed method to characterize the frequency noise of diode lasers foresees the generation of an optical beatnote between two lasers. The frequency of the beatnote ($\nu_b$) is the difference in frequency between the two lasers (named $L1$ and $L2$ respectively), as reported in (\ref{eq:v_beat}).
\begin{equation} 
    \label{eq:v_beat}
    \nu_{b}= \nu_{L1} - \nu_{L2}
\end{equation}
Since the noise of the two independent lasers is uncorrelated, the power spectral density of the beatnote frequency will be the sum of the noise spectra of the two lasers:
\begin{equation}
    S_{\nu_b}(f) = S_{\nu_{L1}}(f)+S_{\nu_{L2}}(f)
    \label{eq:beatnote_PSD}
\end{equation}
where $f$ is the Fourier frequency. As evident from (\ref{eq:v_beat}) and (\ref{eq:beatnote_PSD}), the beatnote is a convenient tool translating the noise of optical oscillators into RF domain. Once the beatnote signal is generated and detected, it can be analyzed by various RF techniques.

Different beatnote schemes can be applied for characterizing the noise of diode lasers: if two lasers are nominally identical, a beatnote can be generated between the two and the measured noise can be scaled by 3 dB, assuming both sources contribute equally to the measured noise. Alternatively, if a lower-noise laser source is available, it can be used as a ``reference'', since the measured noise will be dominated by the other laser (that, in turn, can be considered the ``device under test'').

The beatnote can be generated in different ways, mainly with a fiber coupler or by superimposing the two laser beams in free space on a beam splitter. To maximize the beatnote amplitude, the two beams must be overlapped with the same polarization and spatial mode. Then, the obtained optical-beatnote is acquired by a photodiode. The photodetector converts the produced beatnote from optical to electronic signal. The beatnote signal (in the RF domain) is then converted into a DC voltage using a custom-made frequency-to-voltage converter. The voltage signal in output is finally analyzed using an FFT spectrum analyzer to obtain the power spectral density of the laser frequency noise.

\begin{figure}[t]
\centering
\includegraphics[scale=0.5]{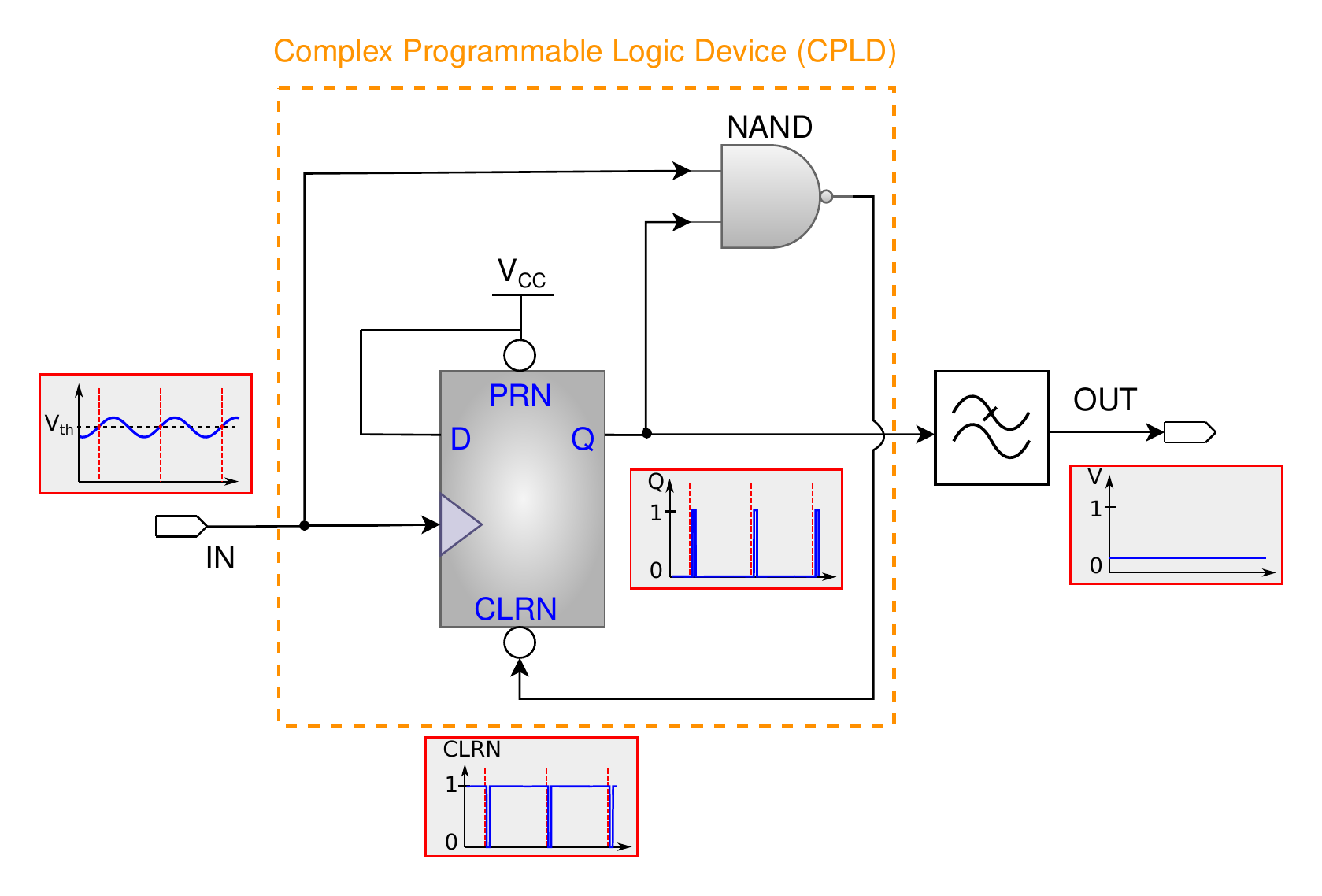}
\caption{f/V converter scheme. The input is biased and sent to one input of a NAND gate and to the trigger of a D-latch (both implemented on a commercial CPLD). The Negated Preset (PRN) and the D-input of the D-latch are both forced to high. The output Q is set to high at each rising edge and quickly reset to zero as soon as the NAND gate sends a ``zero'' at the Negated Clear (CLRN) of the D-latch. The Q output is then averaged through a low-pass filter. The insets show the voltage signals at significant points of the circuit.}
\label{fig:fV}
\end{figure}

\
The f/V converter works with the following principle: it transforms the input RF signal into a train of pulses with fixed amplitude and width. By construction, the pulse-train repetition rate is the same as the input frequency. The pulses are then smoothed out by an analog filter. This results in a DC voltage at the output that will be proportional to the input frequency. In more detail, the device is based on a D-latch and a NAND logic port, configured on a Complex Programmable Logic Device (CPLD). The scheme is shown in \Cref{fig:fV}. The input digital signal from the photodiode is sent both to one port of the NAND and to the D-latch trigger. When a positive edge of the input signal is detected, the output Q of the D-latch follows the D input. The D input is forced to high since it is connected to the power supply. At this point, the NAND detects two signals above the threshold at the input and sends a ''zero`` to the clear input of the D-latch. In this manner, the Q output is reset to ``low'' until the next edge is detected by the trigger. 

\begin{figure}[t]
\centering
\includegraphics[scale=0.5]{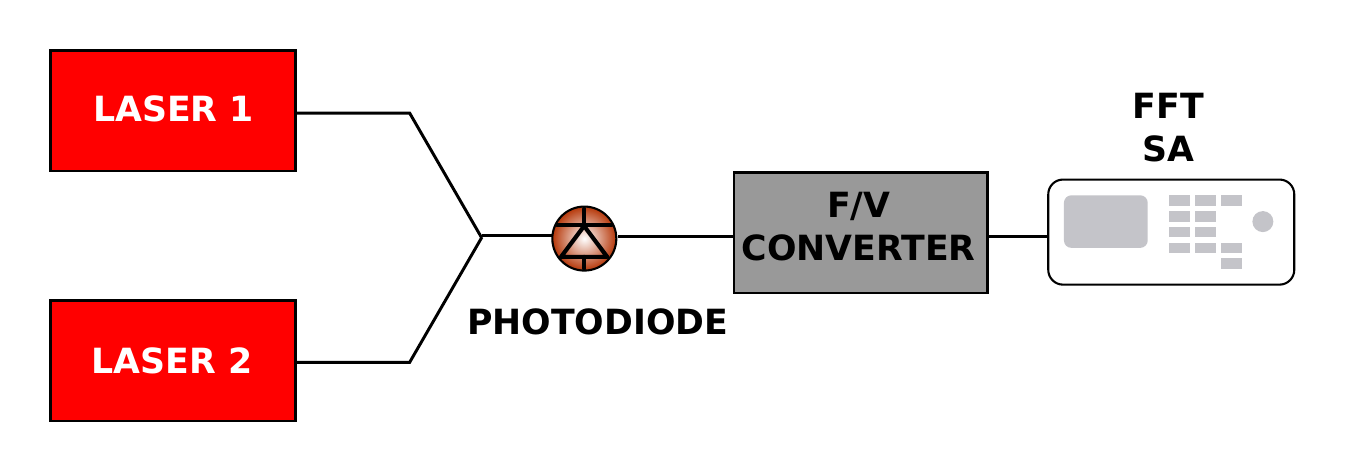}
\caption{Measurement method scheme. An optical beatnote between two lasers is generated (either in free space or with a fiber coupler) and detected by a photodiode. The RF beatnote signal at the output of the photodiode is sent to a frequency-to-voltage converter. The resulting signal is a DC voltage proportional to the beatnote carrier frequency. Knowing the f/V converter sensitivity, an FFT spectrum analyzer (FFT SA) can be used to obtain the PSD of the laser frequency noise.}
\label{fig:scheme}
\end{figure}

The result is a train at very short pulses, whose duration is set by the propagation time of the NAND and the Negated Clear (CLRN) combined. The repetition rate of the pulses follows the input frequency since each pulse is generated at each (positive) zero-crossing of the input signal. The series of pulses arrive at a $\mathrm{3^{rd}}$-order analog low-pass-filter with a cut-off of \SI{1}{\mega\hertz} that averages them, producing an output voltage related to the input frequency simply by a constant multiplicative factor.

As it will be further discussed in \Cref{sec:comparison}, this f/V converter is based on the detection of zero-crossings of the optical beatnote signal, therefore it is insensitive to amplitude noise (AM noise) affecting the laser optical power (and in turn the optical beatnote).
%

The frequency-to-voltage converter is powered by a single $\SI{5}{V}$ supply, followed by a voltage regulator to maximize the power supply rejection ratio. The device accepts input signals with power levels in a range from $\SI{-6}{dBm}$ to $\SI{15}{dBm}$.
The frequency range, by design, is in the range  from \SIrange[]{10}{200}{\mega\hertz} corresponding to an output voltage from $\SI{50.4}{mV}$ to $\SI{1.08}{V}$.

\begin{figure}[t]
\centering
\includegraphics[scale=0.24]{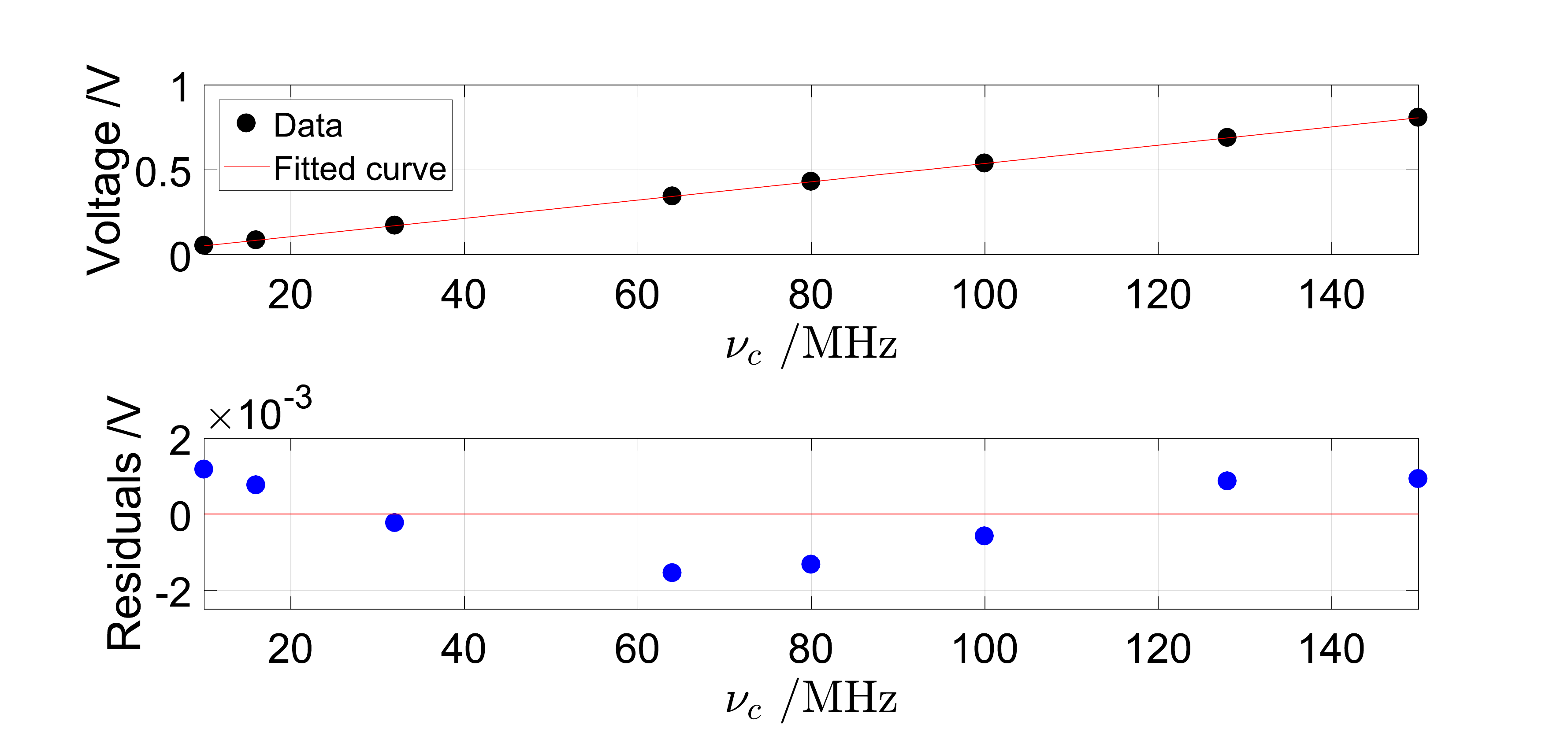}
\caption{Example of calibration data of the f/V converter (upper plot). The experimental data are the voltage measured at the output of the converter when connecting a known input with carrier frequency $\nu_c$. The fitting curve (red) is in the form $y=a+b x$. The lower plot shows the residuals.}
\label{fig:fit_residue}
\end{figure}

\begin{figure}[t]
\centering
\includegraphics[scale=0.24]{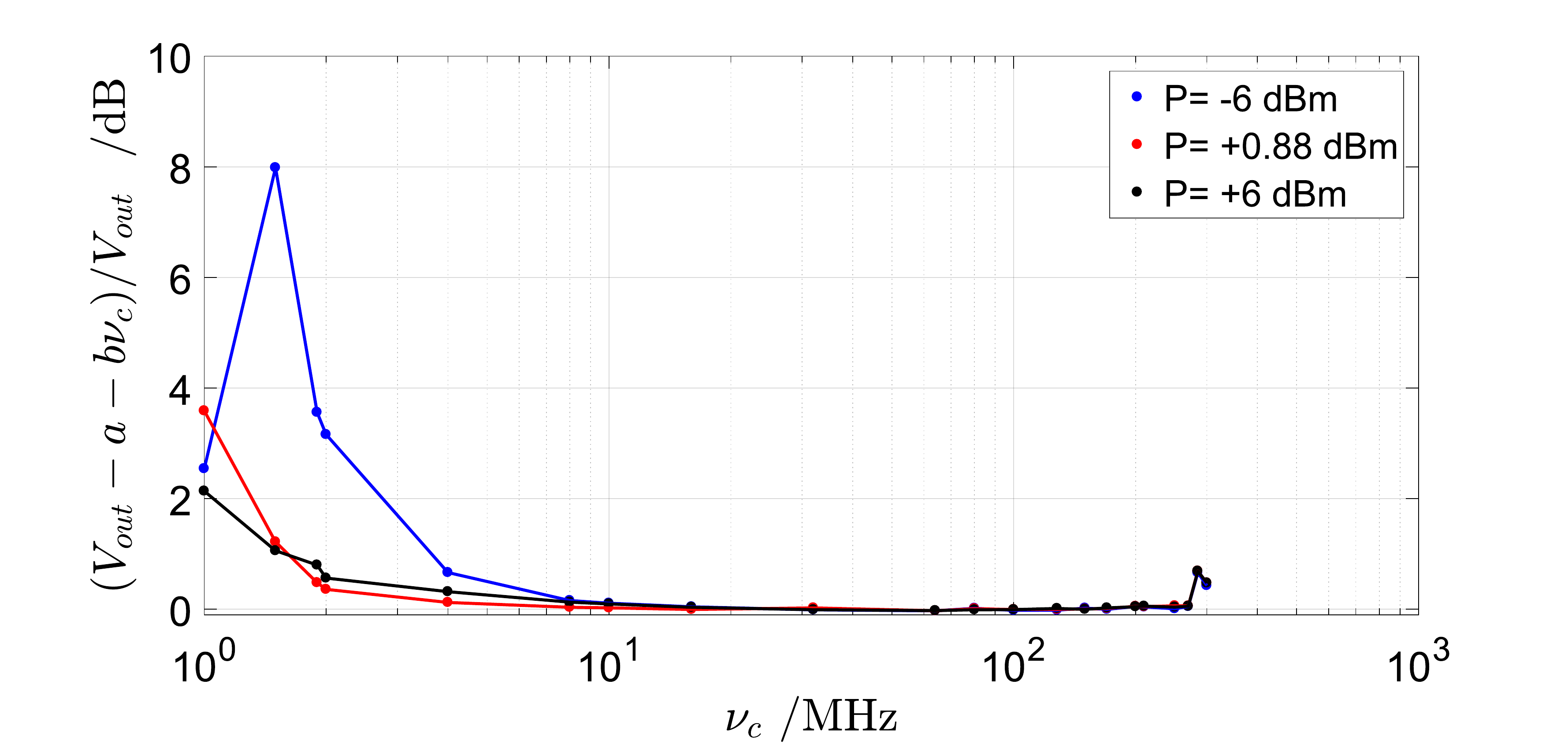}
\caption{f/V converter deviation from linearity at different power input levels, as a function of the carrier frequency $\nu_c$ (in the operational range from $\SI{10}{MHz}$ to $\SI{150}{MHz}$). The y-axis represents the deviation of the obtained values from the linear calibration curve, while the x-axis is the input carrier frequency.}
\label{fig:deviation}
\end{figure}

In order to translate the measured voltage noise into frequency noise, the device needs calibration. For this purpose, sinusoidal signals in the range \SIrange{10}{150}{\mega\hertz}, generated by a RF synthesizer, were given as input to the instrument in place of the beatnote signal. The output of the converter $V_{out}$ was connected to a multimeter to measure the output voltage as a function of the input carrier frequency $\nu_c$. According to the results, a linear behavior is assumed, and a fitting curve $y = a + bx$ is used (\Cref{fig:fit_residue}). In \Cref{fig:deviation} we analyzed the deviation of the output from a linear behavior by calculating the quantity $(V_{out}-a-b \, \nu_c)/V_{out}$ for an extended range of input frequencies. From the latter, we confirmed that in the range from \SIrange{10}{150}{\mega\hertz} the deviation from linearity is less than 0.1 dB, low enough for the device scope. We therefore set it as its operational range. The behavior was reproducible at different input powers.

Since we are primarily interested in noise measurements rather than in measuring the absolute frequency, we can just characterize the instrument with a single sensitivity coefficient $b$. The sensitivity is calculated as ${b=\SI{5.4(1)}{mV/MHz}}$. Since, typically, noise measurements are limited to 0.5 dB uncertainty (due to the limited duration of the acquisition and the front-end electronics), this level of uncertainty on the $b$ coefficient and on the fitting model is adequate.
%
%

\begin{figure}[ht!]
\centering
\includegraphics[scale=0.24]{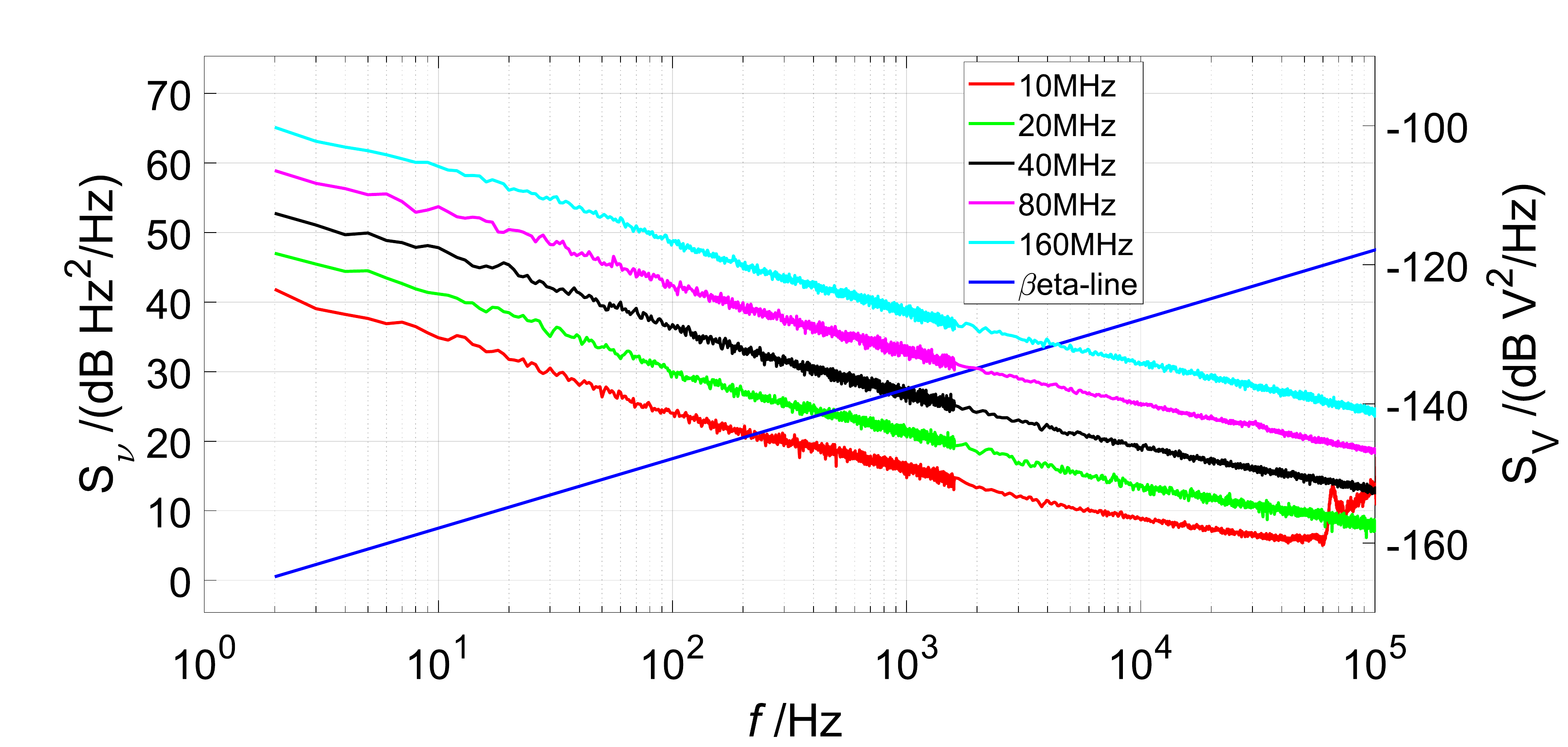}
\caption{Characterization of the instrument noise as a function of the input carrier frequency. The blue line is the function expressed in (\ref{eq:2}) that is used to calculate the contribution of the converter to the measured beatnote linewidth (see \Cref{tab:1}). Right y-axis: f/V converter voltage noise. Left y-axis: f/V converter frequency noise contribution (obtained multiplying the voltage noise by $1/b^2$).}
\label{fig:noise_12dBm}
\end{figure}

To measure the device noise, a low-noise RF signal produced by a synthesizer (Agilent E8257D) is provided as input to the converter. We characterize the instrument noise at different input frequencies, changing the input carrier frequency in octaves from $\SI{10}{MHz}$ to $\SI{160}{MHz}$, at different input levels. A typical measurement of voltage noise is reported in Fig. \ref{fig:noise_12dBm}. 

The obtained values are converted into frequency noise using the calibration coefficient previously determined, with the following expression:
\begin{equation} \label{eq:3}
    S_{\nu}(f)= \frac{1}{b^2} \, S_{V}(f)
\end{equation}
where $ S_{\nu}(f)$ and $ S_{V}(f)$ are the frequency and voltage power spectral densities, respectively.
The f/V noise floor turns out to be flicker, increasing with the carrier frequency about $\SI{6}{dB/oct}$. In the case of beatnote frequency $\nu_b=\SI{80}{MHz}$, at $\SI{1}{kHz}$ the voltage noise is $\SI{-133}{dB \, V^2/Hz}$, which corresponds to $\SI{32.4}{dB \, Hz^2/Hz}$. No appreciable difference in the noise level was observed within the allowed input-power range.

To have a clearer picture of the type of lasers that can be effectively measured with the proposed f/V converter, we can translate the frequency-noise contribution into a contribution to the measured beatnote linewidth. To achieve this, the linewidths corresponding to the measured spectra are calculated with the {$\beta$}-line method \cite{didomenico2010}. Specifically, the spectra are integrated up to the cut-off frequency defined as the point where the PSD crosses the line expressed by the equation:
\begin{equation} 
    \label{eq:2}
    \beta(f)= 8 \, \ln(2) \, \frac{f}{\pi^2}
\end{equation}

The linewidth is then taken as the square root of such integral. The results are shown in \Cref{tab:1}.\\
Notably, the residual noise of the frequency-to-voltage converter has a contribution to the measured linewidth as low as \SI{10}{\kilo\hertz} which is negligible compared to the usual linewidth of diode lasers such as the one characterized in \Cref{sec:validation}.

\begin{table}[ht!]

\footnotesize
\begin{center}
\caption{Instrument noise floor for the f/V converter at different carrier frequencies. The noise is expressed both in terms of frequency noise (converting voltage noise with the sensitivity coefficient $b$) and in terms of a contribution to the beatnote linewidth (integrating the frequency noise over the Fourier-frequency interval prescribed by the ``$\beta$-line'' method).}
\label{tab:1}
{\renewcommand\arraystretch{1.5}
\begin{tabular}[]{ c  c  c }
			\toprule
\thead{Carrier frequency \\ $\nu_c/\mathrm{MHz}$} & \thead{Noise floor \\ $S_{\nu}(f= \SI{100}{Hz})$} & \thead{Linewidth \\ contribution}\\ \hline
10   & 25 dB $\si{Hz^2/Hz}$ & 370 Hz\\ 
20   & 30 dB $\si{Hz^2/Hz}$ & 795 Hz\\ 
40   & 37 dB $\si{Hz^2/Hz}$ & 1.7 kHz\\ 
80   & 43 dB $\si{Hz^2/Hz}$ & 4.3 kHz\\ 
160  & 49 dB $\si{Hz^2/Hz}$ & 10.2 kHz\\ \bottomrule
\end{tabular}}
\end{center}
\end{table}
\subsection{Techniques comparison}
\label{sec:comparison}
To have a more quantitative picture of the possible advantages of the proposed technique with respect to the already mentioned schemes \cite{llopis2011,Volyanskiy2010,Bartalini2011,Maleki2017}, we compare them in terms of some of the main parameters and applicability. Clearly, one of these schemes can be advantageous depending on the laser type, but we can highlight the common features and differences. The comparison is shown in \Cref{tab:comparison}. In the first row, we compare the sensitivity of such schemes to detect frequency fluctuations of the laser under investigation. For most schemes, the sensitivity depends on some experimental parameters, such as the voltage ($V_{DC}$) resulting from the detection of the optical power ($P_{opt}$) or the contrast of the atomic line ($C$). A rough preliminary calibration is therefore needed before each measurement. In the proposed method, the sensitivity is ``hardware-fixed'' instead.

Regarding the achievable bandwidth, the f/V converter performs well, since it can reach the MHz level. It must be noted that in the other schemes, the optical signal does not discriminate frequency noise from amplitude noise ($V_{DC}\propto P_{opt}$), thus the bandwidth at high Fourier frequencies can be effectively limited by amplitude noise (AM) emerging and limiting the measurement sensitivity \cite{Bartalini2011}. The f/V converter is instead immune to AM noise at all Fourier frequencies. 

Of course the main drawback of the proposed technique is the need for an additional laser source at the same wavelength for generating the optical beatnote.

In terms of laser wavelength, self-heterodyne is limited to telecom or near-infrared regions, due to the high losses in the fiber at shorter wavelengths \cite{Willner2020}. The side-of-resonance and DPSK methods need a rather fine wavelength tuning to the set-point of maximum sensitivity.  The acquisition of the beatnote with the f/V converter instead is rather independent of the laser wavelength and does not need tuning of the frequency setpoint. What is important is that the frequency difference between the two lasers lies in the acceptable input range (below \SI{150}{\mega\hertz}).

Finally, all methods are suitable both for measuring the laser with the frequency either free-running or stabilized to an external reference. Of course, in the side-of-resonance or DPSK methods, the free-running frequency must be stable enough to remain close to the point of maximum sensitivity. The heterodyne or beatnote schemes are instead more immune to possible frequency drifts. In the case of the side-of-resonance method, if the stabilization is performed on the top of the same atomic transition (at $\nu=\nu_a)$, such stabilization must be done with an additional modulated optical branch, frequency shifted by half resonance width ($\pm\Delta\nu_a/2$).

\begin{table*}[t] 

\caption{Comparison of the main techniques for measuring the frequency noise of diode laser in terms of sensitivity, applicability and main experimental parameters. $V_{DC}$ is voltage output resulting from the optical detection, $\tau$ the delay of the optical path, $C$ is the atomic resonance contrast, $\Delta\nu_a$ is the atomic resonance linewidth, $1/(2\pi \tau_{LP})$ the bandwidth of the f/V converter low-pass filter. $V_{fs}$ is the ``full-scale'' voltage at the output of the converter expressed in Volt.}
\begin{center}

\label{tab:comparison}
{\renewcommand\arraystretch{2.2}

\begin{tabular}{ c  c  c  c  c  }
\toprule
 
& 
\thead{Self-\\ Heterodyne \\ \cite{llopis2011}} & 
\thead{Side-of-\\resonance \\ \cite{Bartalini2011}} & 
\thead{DPSK \\ \cite{Volyanskiy2010} } & 
\thead{Optical beatnote + \\ f/V converter \\ (this work)  }           \\
\hline
Sensitivity & 
$ \dfrac{V_{DC}}{\sqrt{2}}  2 \pi \tau  \dfrac{\sin(\pi f \tau)}{(\pi f \tau)}$ & 
$\simeq \dfrac{C}{2 \Delta \nu_a}$  & 
$ 2 \pi\tau V_{DC  }$ & 
$\simeq \dfrac{ V_{fs}}{100}\si{\per\mega\hertz} $\\ 
\thead{Bandwidth limit \\(Fourier frequency)} & 
$\simeq 1/\tau$ &
$\simeq \Delta \nu_a$  & 
\thead{detector \\ bandwidth}  & 
$\simeq \dfrac{1}{2 \pi \tau_{LP}} $\\ 
\thead{Reference \\laser} & no & no & no & yes \\ 
\thead{Laser \\ frequency} & 
\thead{Telecom \\ $(\lambda\geq 1200 \;\mathrm{nm})$}  & $\nu_a$ &  
\thead{$(2k+1)/4\tau $ \\ $k \in \mathbb{Z}$  } & any \\ 
\thead{frequency stabilized/ \\ free-running (CL/OL)} & 
OL, CL & 
OL, $\mathrm{CL^*}$&
OL, CL &
OL, CL \\ \bottomrule
\multicolumn{4}{l}{\footnotesize $^*$ Only with additional optical branch frequency-shifted by $\Delta\nu_a/2$.} \\
\end{tabular}}
\end{center}
\end{table*}
\section{Measurement of a DFB laser frequency spectrum}
\label{sec:validation}
\begin{figure}[ht]
\centering
\includegraphics[scale=0.6]{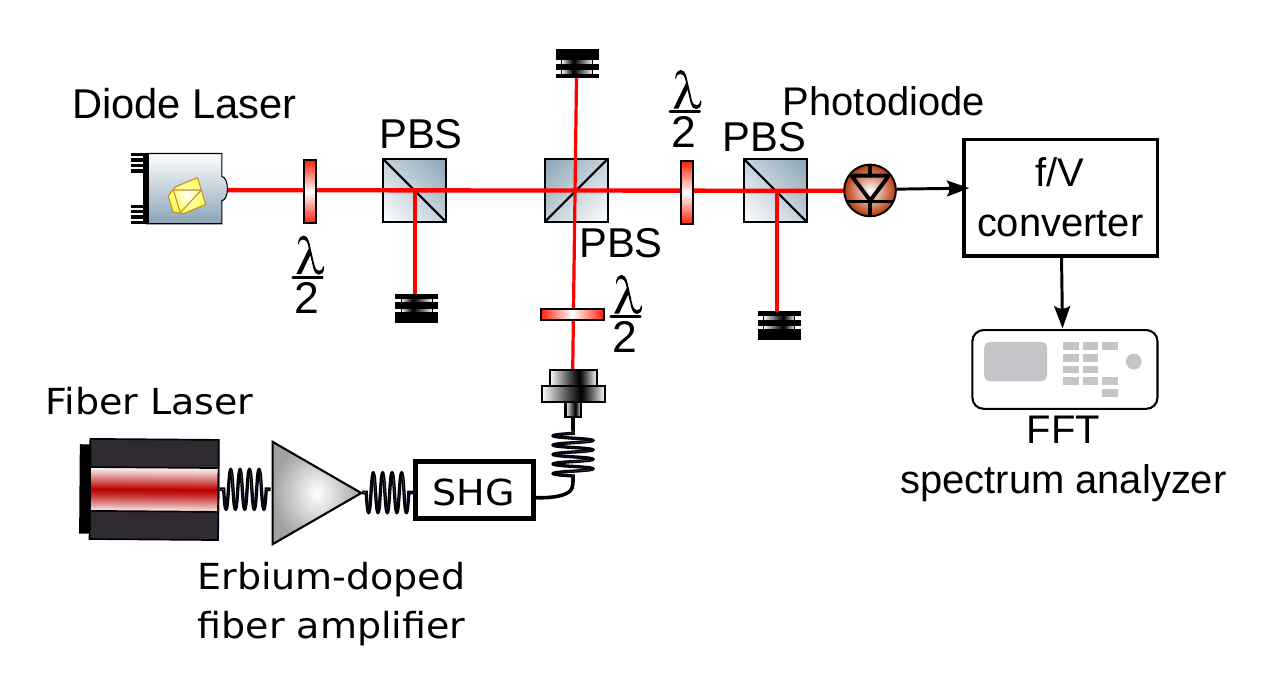}
\caption{Laser setup for beatnote generation. The fiber laser is collimated after the second-harmonic-generation module (SHG) and superimposed with the free-space DFB laser under test.
Half-wavelength ($\lambda/2$) plates and polarizing beam splitters (PBS) are used to overlap the beams, adjust the power level on the two branches, and make the two beams interfere (with the same parallel polarization) before the photodiode. Then, the beatnote is detected by the photodetector. Finally, the detector output is sent to the f/V converter, and the converter output is analyzed by an FFT spectrum analyzer.}
\label{fig:beat_scheme}
\end{figure}

In this section, the measurement technique explained in \Cref{sec:convertitore} is applied to a particular case of study.  
We measured the frequency noise of a distributed-feedback semiconductor laser currently employed in a Rb clock experiment \cite{micalizio2012}. \\
A fiber laser with narrower linewidth is used as a reference laser in the beatnote scheme as shown in Fig. \ref{fig:beat_scheme}. The beatnote between the DFB (``device under test'') and the fiber laser (``reference'') is generated in free space with the use of polarizing optics. The first two polarizing beam splitters (PBS) are used to regulate the optical power on the two branches and to spatially overlap the beams. The last half-wavelength plate and PBS are used to send the two beams onto the photodetector with the same polarization. 
The reference laser is a low-noise single-frequency (Koheras ADJUSTIK by NKT Photonics), co-doped Erbium/Ytterbium fiber laser at $\SI{1560}{nm}$. It is used in combination with a continuous-wave high-power fiber amplifier (Koheras Boostik). A second-harmonic generation crystal doubles the laser frequency to be resonant with the Rb $\mathrm{D_2}$ line at $\SI{780}{nm}$ \cite{almat2018}.
The DFB laser under test and the reference laser are superimposed in free space with the same polarization and power on a dedicated optical bench. The frequency detuning between the two lasers is set close to \SI{80}{\mega\hertz}. The beatnote signal is finally acquired by a Si photodiode (FPD 510-FV), having a bandwidth of $\SI{250}{MHz}$. The photodetector converts the beatnote produced from optic to electronic signal, with a gain of \SI[per-mode=symbol]{4e4}{\volt\per\watt}. Finally, the photodiode signal is sent to the f/V converter, and the converter output is analyzed with an FFT spectrum analyzer (Keysight 35670A).

The frequency noise is measured with the fiber laser and the diode laser free-running. This is the best condition for characterizing the intrinsic properties of the laser diode, without the complication of knowing the control-loop transfer function. Moreover, it provides knowledge of the needed bandwidth and characteristics of an eventual frequency-stabilization loop. We stress that this measurement is easily performed with the f/V converter, even if the laser (and hence the beatnote) frequency fluctuates by several \si{\mega\hertz} over time, since, as reported in \cref{sec:convertitore}, the converter has a linear response over a wide input frequency range.

\begin{figure}[ht]
\centering
\includegraphics[scale=0.24]{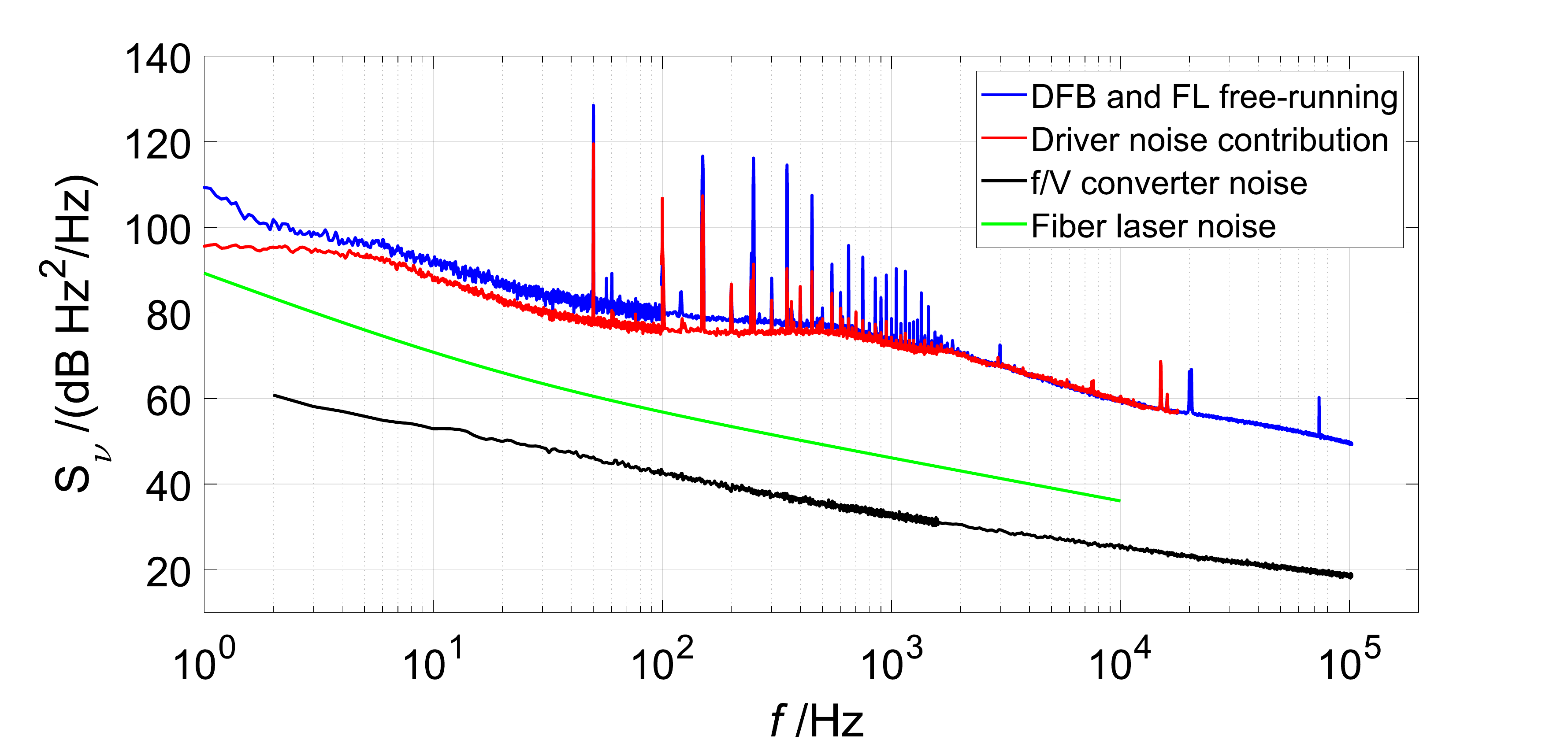}
\caption{Beatnote frequency noise contribution with the DFB and fiber laser in free-running (blue) and current-driver noise contribution (red). The contributions of the f/V converter noise (black) and the reference fiber laser (FL, green) are also reported and they are at least 20 dB lower than the beatnote frequency noise.}
\label{fig:freq_noise}
\end{figure}

The measured noise spectrum of the DFB laser has roughly a flicker-frequency slope ($1/f$) as reported in Fig. \ref{fig:freq_noise}. The linewidth of the diode laser is measured using the $\beta$-line method and results equal to about $\SI{2}{MHz}$, much larger than the one of the reference laser, declared around $\SI{80}{kHz}$. 
The f/V-converter residual noise and the fiber laser noise are also represented in Fig. \ref{fig:freq_noise}. The fiber-laser noise is taken from the test report given by the manufacturer. Both contributions are at least $\SI{20}{dB}$ below the measured diode laser frequency noise. Therefore, the measurement in blue reported in Fig. \ref{fig:freq_noise} represents only the DFB frequency noise.\\
As a further characterization, we analyzed how the current driver impacts the laser performance, as reported in red in Fig. \ref{fig:freq_noise}. The driver current noise contribution is measured independently and converted into frequency noise through the diode current-to-frequency sensitivity coefficient $\frac{\partial \nu_L}{\partial I}$. Thus, the driver noise contribution $S_{\nu, driver}(f)$ is estimated as:
\begin{equation} \label{eqn:3}
   S_{\nu, driver}(f) = \left| \frac{\partial \nu}{\partial I} \right| ^{2} \cdot S_I(f)
\end{equation}

The current-to-frequency coefficient of the DFB laser is derived by measuring the current setpoint when the laser is locked on different Rb transition lines. Knowing the frequency difference between the transitions from the Rb level scheme, we are thus able to extract the coefficient $\partial\nu_L/\partial I$. The accuracy of the optical frequency difference using sub-Doppler spectroscopy is conservatively of the order of \SI{50}{\mega\hertz} (mainly limited by non-linearity in the current-to-frequency relationship). 
In the case of $F=1 \rightarrow F'=1,2$ and $F=2 \rightarrow F'=2,3$  crossover lines, the frequency difference is  \SI{7.050(50)}{\giga\hertz} and the currents setpoints are $I= \SI{95.3}{\milli\ampere}$ and $I= \SI{86.1}{\milli\ampere}$ respectively.
Thus, the experimentally current-to-frequency coefficient is evaluated as:
\\
\begin{equation}
    \frac{\partial \nu_L}{\partial I}=  \frac{\SI{7.050(50)}{\giga\hertz}} { \SI{9.2}{\milli\ampere} } = \SI{766(5)}{\mega\hertz\per\milli\ampere}
\end{equation}
\\
\
This value is used to calculate the current driver contribution reported in Fig. \ref{fig:freq_noise}. From this analysis it is found that the current driver noise contribution is not negligible, limiting the performances of the diode laser, especially for Fourier frequencies above $\SI{500}{Hz}$ of the noise spectrum. In the range from $\SI{10}{Hz}$ to $\SI{500}{Hz}$ the noise induced by the current driver is about $\SI{3}{dB}$ lower than the diode laser noise. A laser driver with a lower frequency noise contribution can be used to improve the diode laser performance and to obtain better knowledge of the intrinsic laser frequency noise. 
%
%

\section{Conclusions}
\label{sec:conclusioni}
In this work we presented a robust and cost-effective method to measure the frequency noise spectrum close to the carrier of semiconductor diode lasers. The method relies on the acquisition of the beatnote between two laser sources, the second one being nominally identical to the laser under test or lower noise, acting as a reference. The beatnote frequency noise is converted into voltage noise by a digital frequency-to-voltage converter and is easily analyzed with FFT algorithms. 

The frequency-to-voltage converter was characterized in terms of residual noise, demonstrating to be suitable for the measurements of lasers having a linewidth as low as 10 kHz. 

We applied the method to measure the frequency-noise PSD of a commercial DFB laser in the Fourier spectral range from $\SI{1}{Hz}$ to $\SI{100}{kHz}$, using a narrower-linewidth fiber laser as a reference.
The measurement method was validated, since the electronic noise floor is at least 20 dB lower than the measured noise, for this class of lasers. It is a robust method, requiring only a one-time simple calibration at the \SI{10}{\percent} level of the f/V converter sensitivity. Although it requires a second laser to generate the optical beatnote, differently from other techniques, the setup calibration does not depend on the laser parameters, such as output power, temperature and wavelength.

This method is rather cost-effective for characterizing low-cost laser diodes since it involves standard optics for the generation of the beatnote and usual laboratory instrumentation (such as FFT analyzers) to perform the measurement. As a matter of fact, the cost of the f/V converter itself is less than one hundred euros.

The setup can be expanded for measuring multiple beatnotes at once, exploiting usual cross-correlation techniques \cite{Vernotte2016}, once the voltage signals have been digitally acquired.
In this way, an absolute characterization of the frequency-noise properties of the individual diode laser sources can be obtained, even in absence of a low-noise reference laser. 

The frequency characterization of diode lasers will support the field of vapor-cell clocks, and will be extremely important in the next few years for developing next-generation chip-scale atomic devices \cite{Kitching2018}. This class of atomic sensors is already having a huge impact on research, technology and industry \cite{Zhan2016,Vicarini2018,Fernandez2017}, and detailed knowledge of the laser source is a step towards performance improvement also in this field.

\section{Acknowledgments} \label{sec:acknowledgments}
The authors would like to thank Elio Bertacco for his invaluable help. This work was partially funded by the European Space Agency under the project GSTP Element 2 Make Industrialization of the Rb POP Clock.

%
\bibliographystyle{IEEEtran}
\bibliography{biblio}

\section{Biography Section}
\vspace{-1cm}
\begin{IEEEbiographynophoto}{G. Antona} was born in Licata, Italy in 1996. He received the Master’s Degree in Electronic Engineering from Polytechnic of Turin, Italy, in 2021 with the thesis "Laser frequency characterization for new generation atomic clocks", developed in the Time and Frequency group of Istituto Nazionale di Ricerca Metrologica (INRIM), Turin, Italy. He joined as speaker the 2022 Conference of European Frequency and Time Forum $\&$ The IEEE International Frequency Control Symposium. Actually he is working in Rete Ferroviaria Italiana in the Technical Department of National ERTMS Plan for the group of programming and monitoring ERTMS Deployment Plan on RFI network. 
\end{IEEEbiographynophoto}
\vspace{-1cm}
\begin{IEEEbiographynophoto}{M. Gozzelino} received his Master’s degree in Physics from the University of Florence, Italy, in 2016. The same year he joined the Time and Frequency group of Istituto Nazionale di Ricerca Metrologica (INRIM), Turin, Italy. In 2018, he was visiting Ph.D. at ICFO (Barcelona, Spain), performing studies on a squeezed laser source. He received the Ph.D. in metrology from Politecnico di Torino in 2020 and is now working as a researcher at INRIM. His research interests include atomic spectroscopy, laser physics and atomic clocks.
\end{IEEEbiographynophoto}
\vspace{-1cm}
\begin{IEEEbiographynophoto}{S. Micalizio} received the Master Degree in Physics from the University of Turin. In 2002, he received a Ph.D. in Metrology from the Politecnico of Turin. His research activity at INRIM is mainly devoted to vapor-cell clocks. In particular, he contributed to the development of a maser based on CPT and of a pulsed optically pumped clock. He was responsible of several contracts funded by the Italian Space Agency, by the European Space Agency and by the European Metrological Research Program.
\end{IEEEbiographynophoto}
\vspace{-1cm}
\begin{IEEEbiographynophoto}{C. E. Calosso} received the Master’s Degree in Engineering from Polytechnic of Turin, Italy, in 1998. In 2001, he was a guest researcher at NIST for studies on multi-launch atomic fountain. In 2002, he received a Ph.D. in communication and electronic engineering from the Polytechnic of Turin. His research activity was devoted to the development of the electronics for the atomic fountain and for the CPT maser. He is now with the INRIM.
\end{IEEEbiographynophoto}
\vspace{-1cm}
\begin{IEEEbiographynophoto}{G. A. Costanzo} was born in S. Benedetto del Tronto, Italy in 1964. He graduated in electronic engineering from the University of Ancona, Italy in 1989. In 1995, he received the Ph.D. degree in metrology from Politecnico di Torino with a dissertation on the development and evaluation of the high-C-field Cs beam frequency standard. He then went to NRLM (Tsukuba, Japan) as a postdoctoral fellow, to work on the realization of the laser system necessary for the Cs fountain. His main interest is on frequency metrology. Actually he is associate professor of Measurements Systems and Sensors at the Politecnico di Torino.
\end{IEEEbiographynophoto}
\vspace{-1cm}
\begin{IEEEbiographynophoto}{F. Levi} received the Master’s Degree in Physics from the University of Torino in 1992. In 1996, he received the Ph.D. in metrology at the Politecnico di Torino. Since 1995 he is with INRIM where, now, he is a research director leading the time and frequency activities. His main research activities are in the field of microwave and optical atomic frequency standards. In 1998 and then  again in 2004-2005, he was guest researcher at NIST t\&f division to study application of laser-cooling technique to atomic frequency standards.
\end{IEEEbiographynophoto}
\end{document}